\documentclass[aps,reprint,amsmath,amssymb,superscriptaddress]{revtex4-2}
\usepackage{graphicx}% Include figure files
\usepackage{dcolumn}% Align table columns on decimal point
\usepackage{bm}% bold math
\usepackage{subfigure}
\usepackage{float}
\usepackage{color}
\usepackage{mathrsfs}
\usepackage{amstext}
\usepackage{booktabs}
\usepackage{siunitx}
\usepackage{lineno}
\begin{document}

% Use the \preprint command to place your local institutional report
% number in the upper righthand corner of the title page in preprint mode.
% Multiple \preprint commands are allowed.
% Use the 'preprintnumbers' class option to override journal defaults
% to display numbers if necessary
% \preprint{}
% \linenumbers\relax

%Title of paper
\title{Nanoradian-Scale Precision in Light Rotation Measurement via Indefinite Quantum Dynamics}

% repeat the \author .. \affiliation  etc. as needed
% \email, \thanks, \homepage, \altaffiliation all apply to the current
% author. Explanatory text should go in the []'s, actual e-mail
% address or url should go in the {}'s for \email and \homepage.
% Please use the appropriate macro foreach each type of information

% \affiliation command applies to all authors since the last
% \affiliation command. The \affiliation command should follow the
% other information
% \affiliation can be followed by \email, \homepage, \thanks as well.
\author{Binke Xia}
\thanks{These authors contributed equally to this work.}
\author{Jingzheng Huang}
\thanks{These authors contributed equally to this work.}
\affiliation{State Key Laboratory of Advanced Optical Communication Systems and Networks, Institute for Quantum Sensing and Information Processing, School of Sensing Science and Engineering, Shanghai Jiao Tong University, Shanghai 200240, China}
\affiliation{Hefei National Laboratory, Hefei 230088, China}
\affiliation{Shanghai Research Center for Quantum Sciences, Shanghai 201315, China}
\author{Hongjing Li}
\author{Zhongyuan Luo}
\author{Guihua Zeng}
\email{ghzeng@sjtu.edu.cn}
\affiliation{State Key Laboratory of Advanced Optical Communication Systems and Networks, Institute for Quantum Sensing and Information Processing, School of Sensing Science and Engineering, Shanghai Jiao Tong University, Shanghai 200240, China}
\affiliation{Hefei National Laboratory, Hefei 230088, China}
\affiliation{Shanghai Research Center for Quantum Sciences, Shanghai 201315, China}

%Collaboration name if desired (requires use of superscriptaddress
%option in \documentclass). \noaffiliation is required (may also be
%used with the \author command).
%\collaboration can be followed by \email, \homepage, \thanks as well.
%\collaboration{}
%\noaffiliation

\date{\today}

\begin{abstract}
The manipulation and metrology of light beams are pivotal for optical science and applications. In particular, achieving ultra-high precision in the measurement of light beam rotations has been a long-standing challenge. Instead of utilizing quantum probes like entangled photons, we address this challenge by incorporating a quantum strategy called "indefinite time direction" into the parameterizing process of quantum parameter estimation. Leveraging this quantum property of the parameterizing dynamics allows us to maximize the utilization of OAM resources for measuring ultra-small angular rotations of beam profile. Notably, a $\SI{}{\nano rad}$-scale precision of light rotation measurement is finally achieved in the experiment, which is the highest precision by far to our best knowledge. Furthermore, this scheme holds promise in various optical applications due to the diverse range of manipulable resources offered by photons.
\end{abstract}

% insert suggested keywords - APS authors don't need to do this
%\keywords{}

%\maketitle must follow title, authors, abstract, and keywords
\maketitle

% body of paper here - Use proper section commands
% References should be done using the \cite, \ref, and \label commands

\section{Introduction}
The rotation of light beams play a vital role in a wide range of optical science and applications, such as optical tweezers\cite{Padgett_2011,David_2019,Stilgoe_2022,Hu_2023}, rotational Doppler effect\cite{PhysRevLett.81.4828,doi:10.1126/science.1239936,Zhang:20}, and light-matter interactions\cite{doi:10.1126/science.1203984,PhysRevLett.109.013901,doi:10.1126/sciadv.aau0981}. Due to the inherent rotational invariance of a conventional Gaussian-profile light beam, it is incapable of measuring angular rotations\cite{Pampaloni_2004}. Traditionally, this challenge has been addressed by harnessing additional properties of photons, including photon entanglement\cite{PhysRevA.83.053829,doi:10.1126/science.1227193,Bouchard:17}, OAM-polarization entanglement\cite{Ambrosio_2013,Cimini_2023}, and discrete spatial modes\cite{Xia:22}. By utilizing these quantum or classical resources associated with photon properties, the current state-of-the-art precision in measuring light rotations has reached the $\SI{}{\micro rad}$ scale\cite{Xia:22}. Nevertheless, there has been limited attention directed towards the dynamics of light's rotation and twist, as well as optimizing the utilization of available resources to enhance the precision in measuring these properties.

In order to advance the ultimate precision in light rotation measurement, we employ the quantum parameter estimation theory and develop an indefinite quantum dynamics strategy for this purpose. Generally, a standard quantum parameter estimation (SQPE) procedure consist of a prepared probe state, a definite dynamical process to encode the parameter onto the probe state, a practical measurement and a classical estimation method to extract the unknown parameter from the final probe state\cite{RevModPhys.89.035002,Liu_2019}. Here, we devise the dynamical evolving involving an indefinite time direction\cite{Chiribella_2022} for encoding the unknown parameter onto the probe state. This indefinite-time-direction quantum parameter estimation (IQPE) scheme employs a two-level auxiliary meter (spin of pseudo-spin state) as a quantum switch, resulting in a superposition of both forward and backward evolving of the system probe. By defining the characteristic operator $\hat{V}_{S}=\partial_{g}\hat{H}_{S}$ for a definite dynamical process described by Hamiltonian $\hat{H}_{S}$, we can evaluate and establish the quantum advantages of our IQPE scheme in comparison to the SQPE scheme concerning the precision limit for parameter estimation. According to the quantum Cramér-Rao (QCR) bound theory\cite{HELSTROM1967101,1054108,Helstrom_1973,DEMKOWICZDOBRZANSKI2015345}, the ultimate precision of estimating parameter $g$ is determined by the quantum Fisher information (QFI)\cite{1976235,Liu_2019,Demkowicz_Dobrza_ski_2020}. In the SQPE scheme, the upper bound of QFI is governed by the maximum fluctuation of operator $\hat{V}_{S}$ ($\propto\Delta\hat{V}_{S}^{2}$)\cite{Pang_2017}. In contrast, in our IQPE scheme, the upper bound of QFI is governed by the maximum amplitude of operator $\hat{V}_{S}$ ($\propto\hat{V}_{S}^{2}$), which encompasses both the fluctuation and the mean energy of the operator ($\langle\hat{V}_{S}^{2}\rangle=\langle\Delta\hat{V}_{S}^{2}\rangle+\langle\hat{V}_{S}\rangle^{2}$). Consequently, we harness the full potential of available resources of the dynamical process for quantum parameter estimation. 

In the realm of light science, there is a wide range of manipulable resources associated with photon properties, including polarization, orbital angular momentum (OAM), and more. Hence, our scheme exhibits significant potential for applications within optical systems. Given that the dynamics of the rotational process is characterized by the OAM operator, the maximum available resources is determined by the OAM values of probe states. Hence, we employ the $N$-order Laguerre-Gaussian (LG) beam with orbital angular momentum (OAM) value $l=N$ as the system probe, while utilizing the polarization state as the auxiliary quantum switch to generate an indefinite-rotation-direction dynamical process for the probe state (LG beam). Traditionally, the pure LG beam is unable to measure the angular rotations as its rotational invariance. Nevertheless, our scheme is successful to measure the rotations with the precision enhanced by factor $N$, which follows the same power law of Heisenberg scaling. Notably, our experimental results finally achieve a precision of $\SI{12.9}{\nano rad}$ in angular rotation measurement using a 150-order LG beam.

\section{Results}
\subsection{Quantum-enhanced precision limit with indefinite time direction}
\begin{figure}[h]
	\centering
	\includegraphics[width=\linewidth]{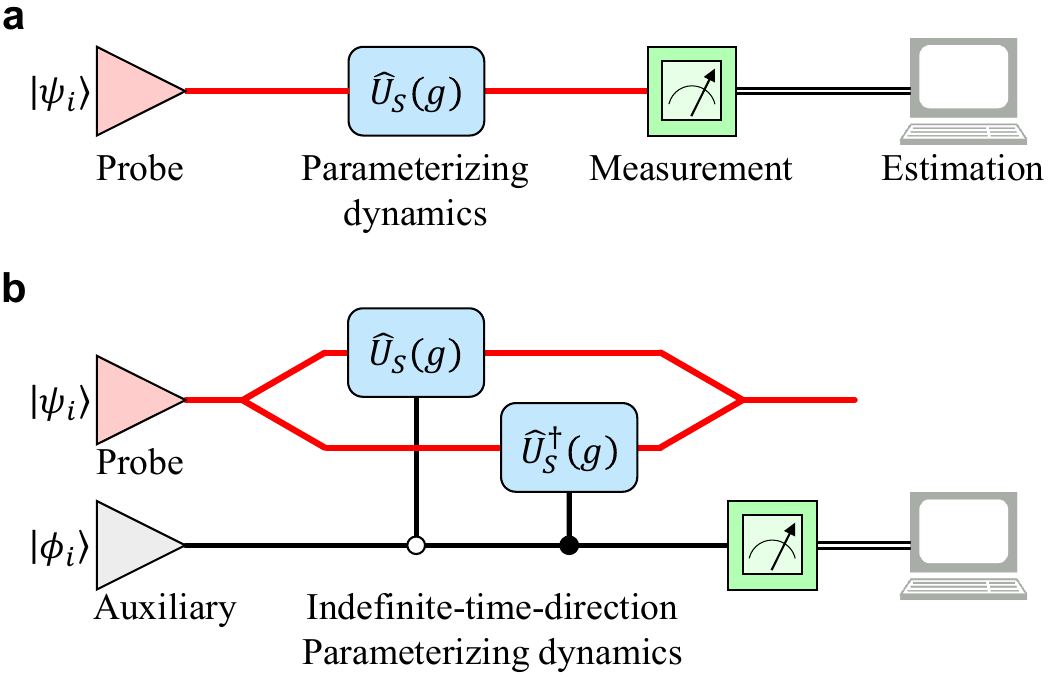}
	\caption{\label{fig:1} Schematic of the quantum parameter estimation procedure. \textbf{\textsf{a}} Standard quantum parameter estimation procedure. \textbf{\textsf{b}} Quantum parameter estimation procedure with indefinite time direction.}
\end{figure}

In a standard quantum parameter estimation (SQPE) procedure depicted in Fig. \ref{fig:1}\textbf{\textsf{a}}, the system probe is initialized in some state $|\psi_{i}\rangle$ and evolves under a parameter-dependent Hamiltonian $\hat{H}_{S}(g)$ with $g$ as the parameter to estimate. After an evolution for some time $T$, we measures the final state $|\psi(g)\rangle=\hat{U}_{S}(0\to T)|\psi_{i}\rangle$, where $\hat{U}_{S}(0\to T)$ is the unitary dynamics under the Hamiltonian $\hat{H}_{S}(g)$ for time $T$. To evaluate the precision with respect to the parameter $g$, we employ the quantum Fisher information (QFI) $\mathcal{Q}(g)$ as a figure of merit\cite{PhysRevLett.96.010401}. Then the quantum Cramér-Rao (QCR) bound $\delta\hat{g}^{2}\ge 1/[\nu\mathcal{Q}(g)]$ gives the quantum limit of precision for the parameter $g$, where $\nu$ is the number of independent identical samples. By defining the generator $\hat{\mathcal{H}}_{S}(g)=\mathrm{i}\hat{U}^{\dagger}_{S}(0\to T)\partial_{g}\hat{U}_{S}(0\to T)$, the QFI in the SQPE scenario can be calculated as $\mathcal{Q}_{S}(g)=4\mathrm{Var}[\hat{\mathcal{H}}_{S}(g)]_{|\psi_{i}\rangle}$. Since the generator can be expressed as $\hat{\mathcal{H}}_{S}(g)=\int_{0}^{T}\hat{U}_{S}^{\dagger}(0\to t)(\partial_{g}\hat{H}_{S})\hat{U}_{S}(0\to t)\mathrm{d}t$\cite{Pang_2017}, the upper bound of QFI $\mathcal{Q}_{S}(g)$ is obtained as:
\begin{equation}
	\label{eq:1}
	\mathcal{Q}_{S}(g) \le \left\{\int_{0}^{T}\left[\lambda_{M}(t)-\lambda_{m}(t)\right]\mathrm{d}t\right\}^{2},
\end{equation}
where $\lambda_{M}(t)$ and $\lambda_{m}(t)$ are the maximum and minimum eigenvalues of operator $\partial_{g}\hat{H}_{S}$. We denote this characteristic operator as $\hat{V}_{S}=\partial_{g}\hat{H}_{S}$, which is solely determined by the Hamiltonian of the parameterizing dynamics.

In this work, we propose an indefinite-time-direction quantum parameter estimation (IQPE) scheme to improve the ultimate precision limit compared to the SQPE procedure. As illustrated in Fig. \ref{fig:1}\textbf{\textsf{b}}, we introduce an two-level auxiliary meter $|\phi_{i}\rangle$ (particle or physical dimension), which is prepared together with the system probe $|\psi_{i}\rangle$. This auxiliary meter serves as a quantum switch for the parameterizing dynamics, leading to a indefinite-time-direction evolving of system probe:
\begin{equation}
	\label{eq:2}
	\hat{U}_{I}(0\to T) = \hat{U}_{S}|0\rangle\langle 0|+\hat{U}_{S}^{\dagger}|1\rangle\langle 1|,
\end{equation}
leads to the generator in the IQPE scheme being represented as $\hat{\mathcal{H}}_{I}(g)=\hat{\mathcal{H}}_{S}|0\rangle\langle 0|-\hat{U}_{S}\hat{\mathcal{H}}_{S}\hat{U}_{S}^{\dagger}|1\rangle\langle 1|$. The auxiliary meter is chosen as the maximum superposition state $|\phi_{i}\rangle=\frac{1}{\sqrt{2}}(|0\rangle+|1\rangle)$, then the upper bound of QFI in this IQPE scenario can be calculated as
\begin{align}
	\label{eq:3}
	\mathcal{Q}_{I}(g) &= 4\mathrm{Var}\left[\hat{\mathcal{H}}_{I}(g)\right]_{|\psi_{i}\rangle|\phi_{i}\rangle} \nonumber \\
	&\le \max\left\{\left[\int_{0}^{T}2\lambda_{M}(t)\mathrm{d}t\right]^{2}, \left[\int_{0}^{T}2\lambda_{m}(t)\mathrm{d}t\right]^{2}\right\}.
\end{align}

Comparing the QFIs in the SQPE and IQPE schemes, we can conclude that
\begin{equation}
	\label{eq:4}
	\sup[\mathcal{Q}_{S}(g)] \preceq \sup[\mathcal{Q}_{I}(g)],
\end{equation}
as the upper bound of $\mathcal{Q}_{S}(g)$ is determined by the maximum fluctuation of the characteristic operator $\hat{V}_{S}$, while the  the upper bound of $\mathcal{Q}_{I}(g)$ is determined by the maximum amplitude of $\hat{V}_{S}$.

To demonstrate the superior performance of our IQPE scheme in terms of precision limit enhancement compared to the SQPE scheme in detail, we focus on a widely encountered quantum metrological scenario. In this scenario, the parameter $g$ is encoded linearly onto the probe through a time-independent Hamiltonian $\hat{H}_{S}=g\hat{V}_{S}$. By normalizing the evolving time as $T=1$, the parameterizing dynamics is described as $\hat{U}_{S}(g)=\exp(-\mathrm{i}g\hat{V}_{S})$. In this case, the QFIs of the SQPE and IQPE schemes can be calculated as $\mathcal{Q}_{S}(g)=4(\langle\psi_{i}|\hat{V}_{S}^{2}|\psi_{i}\rangle-\langle\psi_{i}|\hat{V}_{S}|\psi_{i}\rangle^{2})=4\langle\Delta\hat{V}_{S}^{2}\rangle_{i}$ and $\mathcal{Q}_{I}(g)=4\langle\psi_{i}|\hat{V}_{S}^{2}|\psi_{i}\rangle=4\langle\hat{V}_{S}^{2}\rangle_{i}$, respectively. According to the QCR bound theory, the quantum limits of estimating precision of parameter $g$ is governed by the uncertainty relations
\begin{align}
	\text{SQPE:}\quad&\delta\hat{g}^{2}\langle\Delta\hat{V}_{S}^{2}\rangle_{i} \ge \frac{1}{4\nu}, \label{eq:5} \\
	\text{IQPE:}\quad&\delta\hat{g}^{2}\langle\hat{V}_{S}^{2}\rangle_{i} \ge \frac{1}{4\nu}, \label{eq:6}
\end{align}
for the SQPE and IQPE schemes, respectively. Comparing to the SQPE scheme, our scheme enables the mean energy of the characteristic operator of the dynamical process for estimating parameter additionally, due to $\langle\hat{V}_{S}^{2}\rangle=\langle\Delta\hat{V}_{S}^{2}\rangle+\langle\hat{V}_{S}\rangle^{2}$.

It is worth noting that the characteristic operator $\hat{V}_{S}$ is various in different quantum metrological scenarios, thus, there exists a wide range of resources associated with various physical quantities that can be utilized for quantum parameter estimation\cite{Cimini_2023}. In the realm of light science, there are diverse available resources including the polarization and OAM of light can be utilized for optical measurements\cite{Ambrosio_2013,PhysRevLett.112.103604,Barboza_2022}. Therefore, our theoretical results are promising for their application in diverse metrological systems that involves various photon properties.

\subsection{IQPE for optical measurements}
To demonstrate the superiority of our scheme in optical systems , we first consider the scenario of measuring the optical Kerr effect induced phase shift\cite{PhysRevLett.114.210801,Chen_2018}, the dynamical process is described by $\hat{U}_{S}=\exp(-\mathrm{i}\theta\hat{n})$, where $\hat{n}$ is the photon number operator. In this case, the characteristic operator $\hat{n}$ of dynamical process is directly associated with the utilized energy resources. Assuming that the system probe is prepared in a coherent state, i.e., $|\psi_{i}\rangle=|\alpha\rangle$. Based on our previous analysis, the QFI of parameter $\theta$ when adopting the SQPE scheme is given by
\begin{equation}
	\label{eq:7}
	\mathcal{Q}_{S}(\theta) = 4\langle\Delta\hat{n}^{2}\rangle_{i} = 4\bar{n},
\end{equation}
which leads to the standard quantum limit $\delta\hat{g}\ge 1/2\sqrt{\nu\bar{n}}$, as only the fluctuation of the probe's energy is utilized. However, it is worth noting that the mean energy of the probe is proportional to $\bar{n}$, which can provide resources that scale quadratically larger than its energy fluctuation. As a result, the QFI of parameter $\theta$ when using our IQPE scheme can be improved as
\begin{equation}
	\label{eq:8}
	\mathcal{Q}_{I}(\theta) = 4\langle\hat{n}^{2}\rangle_{i} = 4\bar{n}^{2}+4\bar{n},
\end{equation}
where a Heisenberg-scaling term $\bar{n}^{2}$ is involved. This result indicates the capability of achieving Heisenberg-scaling limit using only classical probe states within our scheme.

\begin{figure*}[htb]
	\centering
	\includegraphics[width=0.9\linewidth]{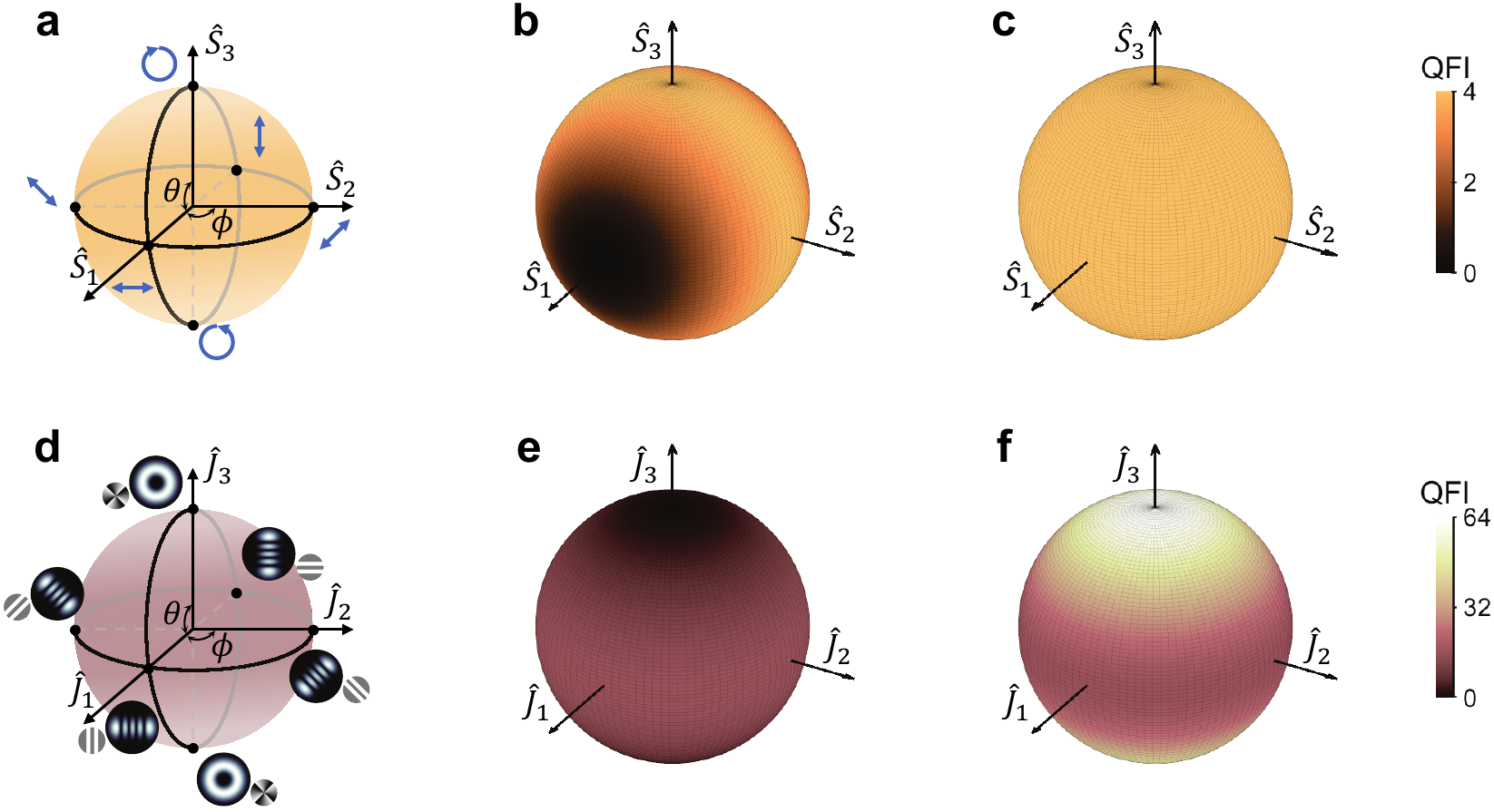}
	\caption{\label{fig:2} Quantum Fisher information of the SQPE and the IQPE schemes. \textbf{\textsf{a}} Classical PS for polarization states. The northern and southern poles represent the right- and left-handed circular polarization states respectively. The states on the equator are linear polarization. \textbf{\textsf{b}} QFI of the birefringent phase $\varphi$ in the SQPE procedure with respect to various polarization states. \textbf{\textsf{c}} QFI of the birefringent phase $\varphi$ in the IQPE procedure with respect to various polarization states. \textbf{\textsf{d}} Modal PS for HLG modes with $N=l=4$. The northern and southern poles represent the LG modes with opposite topological charges. The states on the equator are HG modes. \textbf{\textsf{e}} QFI of the rotation angle $\alpha$ in the SQPE procedure with respect to various HLG beams. \textbf{\textsf{f}} QFI of the rotation angle $\alpha$ in the IQPE procedure with respect to various HLG beams. The values of QFIs are represented by the color intensity in \textbf{\textsf{b}}, \textbf{\textsf{c}}, \textbf{\textsf{e}}, and \textbf{\textsf{f}}.}
\end{figure*}

In order to delve into more practical applications, we explore two optical measurement scenarios: the birefringent phase shift of the polarization states and the angular rotation of the generalized Hermite-Laguerre-Gaussian (HLG) beams. For an arbitrary polarization (spin) state, it can be represented on a classical Poincaré sphere (PS) as shown in Fig. \ref{fig:2}\textbf{\textsf{a}}, where the poles are the right- and left-handed circular polarization states respectively. The axes $\hat{S}_{1}$, $\hat{S}_{2}$ and $\hat{S}_{3}$ represent Stokes parameters, which can be normalized as the Pauli matrices. In this geometric representation, an arbitrary polarization state can be obtained from the right-handed circular polarization state $|R\rangle$ through an Euler rotation $|\psi_{i}\rangle=\exp(-\mathrm{i}\frac{\hat{S}_{3}}{2}\phi)\exp(-\mathrm{i}\frac{\hat{S}_{2}}{2}\theta)|R\rangle$.

The birefringence process of the polarization state is described as $\hat{U}_{S}(\varphi)=|H\rangle\langle H|\exp(-\mathrm{i}\varphi)+|V\rangle\langle V|\exp(-\mathrm{i}\varphi)$, where $|H\rangle$ and $|V\rangle$ represent the horizontal and vertical polarization states, respectively. The parameter $\varphi$ denotes the birefringent phase being measured. Hence, the generator of the birefringence process is defined as $\hat{H}_{S}=|H\rangle\langle H|-|V\rangle\langle V|=\hat{S}_{1}$. Based on our previous theoretical analysis, the QFI of the birefringence parameter $\varphi$ in the SQPE procedure can be calculated as
\begin{equation}
	\label{eq:9}
	\mathcal{Q}_{S}(\varphi) = 4\langle\Delta\hat{S}_{1}^{2}\rangle_{i}= 4-4\sin^{2}(\theta)\cos^{2}(\phi).
\end{equation}
We depict this result in Fig. \ref{fig:2}\textbf{\textsf{b}}, where the QFI of various polarization states is represented by the color intensity on the classical PS. The results show that the polarization states nearby the $\hat{S}_{1}$-axis yield a trivial QFI value though they have the maximum energy associated with the generator $\hat{S}_{1}$. The maximum QFI is only observed on the $\hat{S}_{2}$-$\hat{S}_{3}$ plane, where the states exhibit the maximum fluctuation associated with the generator $\hat{S}_{1}$. This result indicates that there is a precision "dead zone" when employing SQPE scheme. By maximizing the utilization of these two components of resources associated with the generator $\hat{S}_{1}$, our IQPE scheme improves the QFI of the birefringence parameter $\varphi$ as
\begin{equation}
	\label{eq:10}
	\mathcal{Q}_{I}(\varphi) = 4\langle\hat{S}_{1}^{2}\rangle_{i}= 4.
\end{equation}
As shown in Fig. \ref{fig:2}\textbf{\textsf{c}}, the QFI of various polarization states exhibits homogeneity on the classical PS. In comparison to the results of the SQPE scheme, our IQPE scheme also possesses the capability to overcome the precision "dead zone" by incorporating the resource associated with the absolute energy of the generator $\hat{S}_{1}$.

To harness additional valuable resources related to photon properties, we conducted an investigation focusing on the angular rotations of the beam profile. In this scenario, the evolving process is described by $\hat{U}_{S}(\alpha)=\exp(-\mathrm{i}\alpha\hat{L}_{z})$, where the generator $\hat{L}_{z}$ is the OAM operator, and the parameter $\alpha$ is the rotation angle being measured. Therefore, the available resources for parameter estimation in this metrological scenario is associated with the OAM resources of probe states. Here, we employ the HLG beams as the system probes. The analogy between the HLG modes and the quantum eigenstates of a two-dimensional harmonic oscillator (2DHO) allows us to explore the transformation and geometric properties of HLG beams using the Schwinger oscillator model\cite{10.1063/1.1703727,sakurai_napolitano_2017}. Then the mode state of any arbitrary HLG beam can be represented on the modal PS\cite{Padgett:99,Calvo:05,Forbes_2021}, as shown in Fig. \ref{fig:2}\textbf{\textsf{d}}, where the axes $\hat{J}_{1}$, $\hat{J}_{2}$ and $\hat{J}_{3}$ are the angular momentum operators in the oscillator model. The poles represent the LG modes with opposite OAM values. The LG mode at the northern pole can be denoted by the ket $|N,l\rangle$, where the order $N$ determines the Gouy Phase, $l$ is the OAM value (which is also called topological charge). Then an arbitrary HLG beam state of the same order can be obtained from LG beam $|N,l\rangle$ on the modal PS through an Euler rotation $|\psi_{i}\rangle=\exp(-\mathrm{i}\hat{J}_{3}\phi)\exp(-\mathrm{i}\hat{J}_{2}\theta)|N,l\rangle$. (The detailed geometric representation of HLG beams is provided in the method part.)

Here, we investigate the $N$-order HLG beams with the maximum OAM value $l=N$ in the angular rotation process, the QFI of the rotation parameter $\alpha$ within the SQPE procedure and the IQPE procedure can be calculated as
\begin{align}
	\mathcal{Q}_{S}(\alpha) &= 4\langle\Delta\hat{L}_{z}^{2}\rangle_{i} = 4N\sin^{2}(\theta), \label{eq:11} \\
	\mathcal{Q}_{I}(\alpha) &= 4\langle\hat{L}_{z}^{2}\rangle_{i} = 4N^{2}\cos^{2}(\theta)+4N\sin^{2}(\theta), \label{eq:12}
\end{align}
respectively. Notably, the QFI in our IQPE scheme involves a term proportional to $N^{2}$, which exhibits the same power law of Heisenberg scaling when considering the resources associated with OAM. Fig. \ref{fig:2}\textbf{\textsf{d}} depicts the modal PS for HLG beams with the order of $N=l=4$. The corresponding QFIs of rotation parameter $\alpha$ in the SQPE procedure and the IQPE procedure are illustrated in Fig. \ref{fig:2}\textbf{\textsf{e}} and Fig. \ref{fig:2}\textbf{\textsf{f}}, respectively. Furthermore, our scheme also possesses the capability to overcome the precision "dead zone" in this metrological scenario. In the SQPE scheme, states near the $\hat{J}_{3}$-axis result in a trivial QFI value. However, in our IQPE scheme, an additional term of QFI proportional to $N^{2}\cos^{2}(\theta)$ is introduced, effectively eliminating the precision "dead zone", as shown in Fig. \ref{fig:2}\textbf{\textsf{f}}.

\subsection{OAM-enhanced rotation measurement}
Practically, we employ the rotation-invariant HLG beams (LG beams) as the system probe for light rotation measurement.
\begin{figure}[b]
	\centering
	\includegraphics[width=0.8\linewidth]{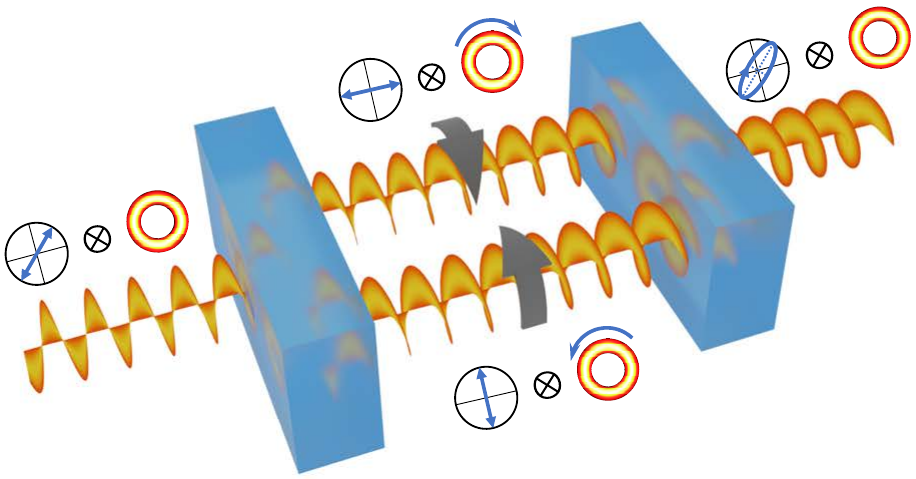}
	\caption{\label{fig:3} Schematic of light rotation process with indefinite time direction.}
\end{figure}
To maximize the available OAM resource, we assign the OAM value of LG beam as its maximum value, i.e., $l=N$. For simplicity, we denote the system probe as the ket $|l\rangle$. Subsequently, we introduce the polarization state as the auxiliary meter to implement the indefinite-time-direction rotation dynamics. In this setup, we assign the orthogonal bases of auxiliary meter as the horizontal polarization state $|H\rangle$ and the vertical polarization state $|V\rangle$. The initial state of auxiliary meter is prepared as the $\ang{45}$ linear polarization state $|+\rangle=\frac{1}{\sqrt{2}}(|H\rangle+|V\rangle)$. The rotation process with indefinite time direction is depicted in Fig. \ref{fig:3}, the auxiliary meter serves as a quantum switch, resulting in a superposition of a pair of opposite rotations on the system probe. This indefinite-time-direction rotation dynamics can be described as (evolving time is normalized as $T=1$):
\begin{equation}
	\label{eq:13}
	\hat{U}_{I}(\alpha) = |H\rangle\langle H|\exp(-\mathrm{i}\alpha\hat{L}_{z})+|V\rangle\langle V|\exp(\mathrm{i}\alpha\hat{L}_{z}).
\end{equation}
Substituting $|\psi_{i}\rangle=|l\rangle$ and $|\phi_{i}\rangle=|+\rangle$, the final state of the entire system after the indefinite-time-direction rotation process can be expressed as
\begin{equation}
	\label{eq:14}
	|\Psi(\alpha)\rangle = \frac{1}{\sqrt{2}}\left(\mathrm{e}^{-\mathrm{i}l\alpha}|H\rangle+\mathrm{e}^{\mathrm{i}l\alpha}|V\rangle\right)|l\rangle.
\end{equation}

Obviously, the profile rotations of the system probe are converted into the polarization changes of the auxiliary meter. This property enables us to design a parameter-independent measurement on the auxiliary meter to extract the rotation parameter $\alpha$. By applying a pair of the projective measurements $\left\{|L\rangle\langle L|, |R\rangle\langle R|\right\}$ on the auxiliary meter, where $|L\rangle$ and $|R\rangle$ represent the left- and right-handed circular polarization states, respectively, we can detect the rotation parameter $\alpha$ with the ultimate precision
\begin{equation}
	\label{eq:15}
	\delta\hat{\alpha}_{\min} = \frac{1}{2l\sqrt{\nu}} = \frac{1}{2N\sqrt{\nu}},
\end{equation}
which reaches a sensitivity following the same power law of Heisenberg scaling at the perspective of utilizing the resources associated with the OAM of the probe state.

\subsection{Experimental scheme of light rotation measurements}
\begin{figure}[hbt]
	\centering
	\includegraphics[width=\linewidth]{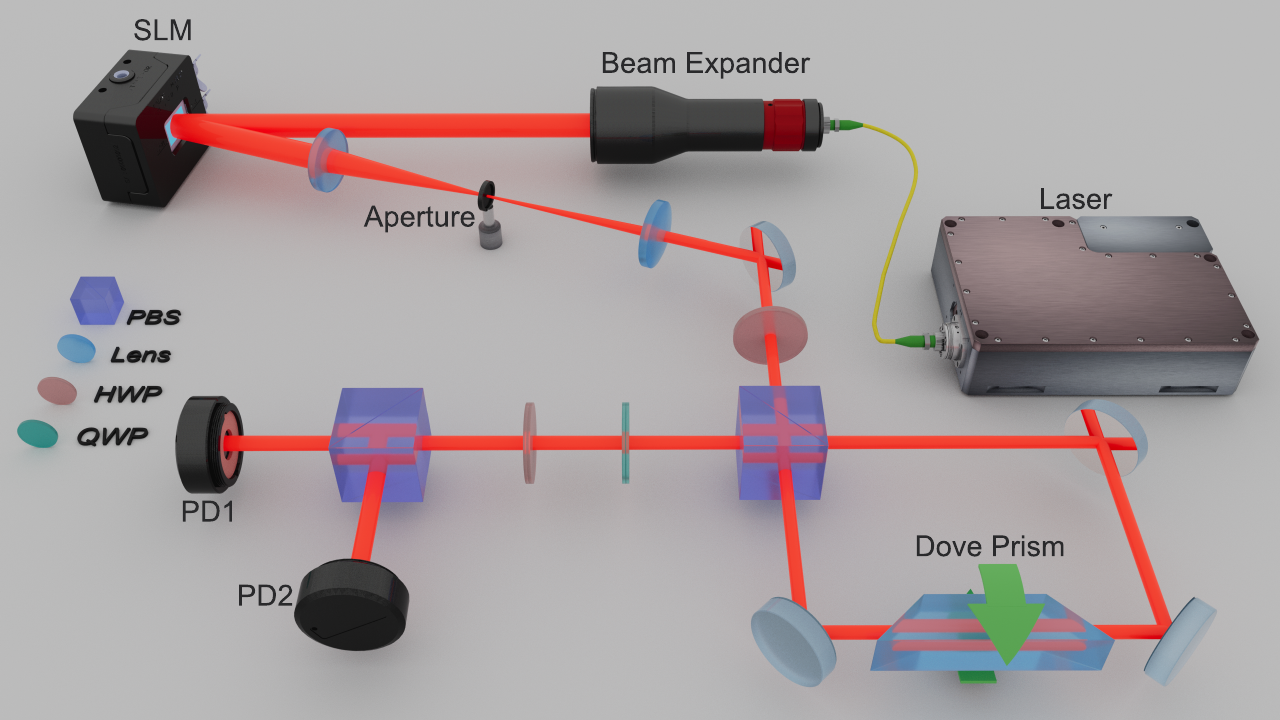}
	\caption{\label{fig:4} Schematic of experimental setup. The $N$-order LG beam with the maximum OAM value $l=N$ is generated using a SLM and a spatial filter system. The polarization state is adjusted to an orientation angle of $\ang{45}$ using a HWP. The indefinite-time-direction rotation process is implemented in a polarized Sagnac interferometer, incorporating a Dove prism to enable the rotation of beam profile. The projective measurements of polarization are performed using a QWP, a HWP, and a PBS. Photodetectors PD1 and PD2 capture the projective photons.}
\end{figure}

Drawing on the superiority of our IQPE scheme, we apply it to the light rotation measurement in a practical optical system. The schematic of experimental setup is illustrated in Fig. \ref{fig:4}, which comprises three main components: preparation of the system probe and the auxiliary meter, implementation of the indefinite-time-direction rotation process, and projective measurements performed on the auxiliary meter.

\begin{figure*}[hbt]
	\centering
	\includegraphics[width=\linewidth]{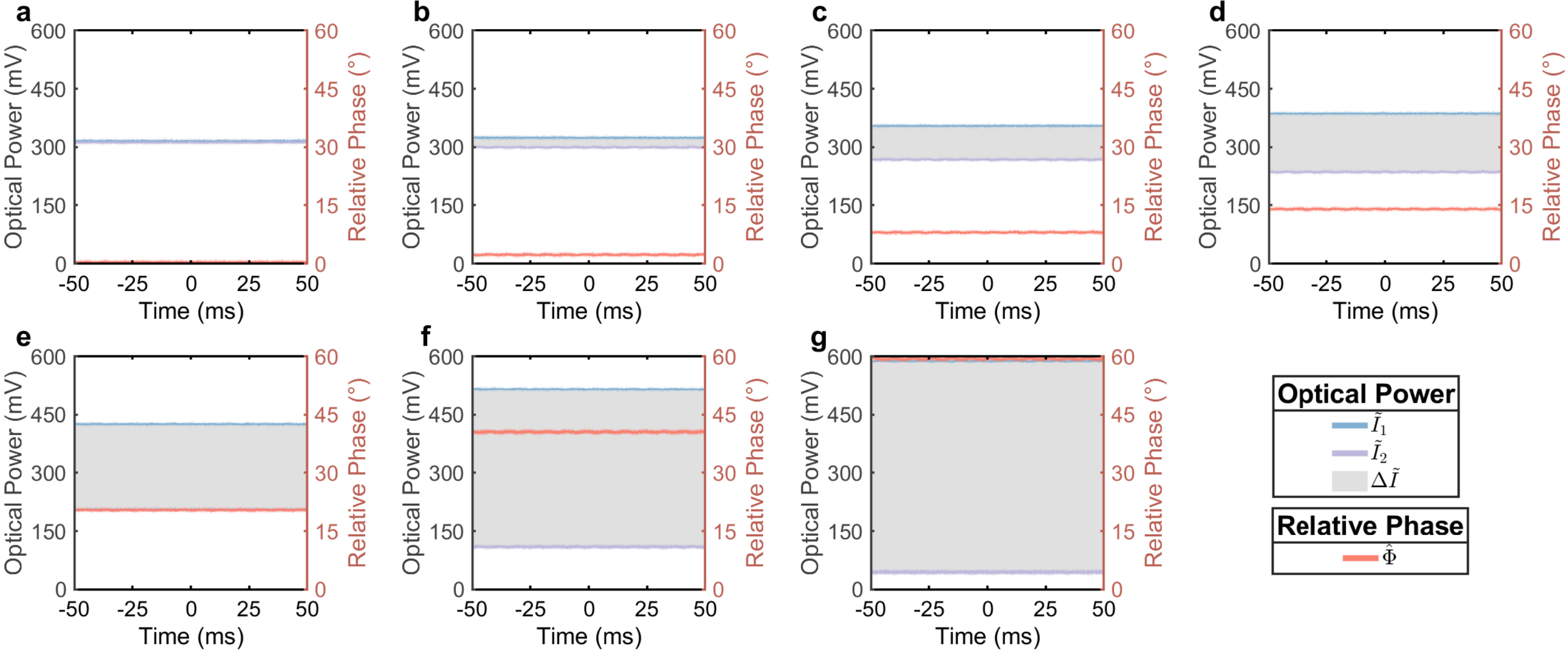}
	\caption{\label{fig:5} Experimental results of measured optical powers and relative phases. The blue lines and the purple lines represent the measured optical powers of PD1 and PD2, respectively. The corresponding values are labeled at the left y-axis. The gray shadows stand for the differential optical powers between two PDs. The orange lines represent the demodulated relative phases. The corresponding values are labeled at the right y-axis. \textbf{\textsf{a}} Experimental results when inputting the Gaussian beam. The relative phase $\Phi$ is solely determined by the additional relative phase $\Delta\varphi$ induced by systemic imperfections. \textbf{\textsf{b}}-\textbf{\textsf{g}} Experimental results when inputting the $N$-order LG beams with OAM values of $l=N=1, 4, 7, 10, 20, 30$. The relative phase $\Phi$ is determined by both the additional relative phase $\Delta\varphi$ and the angular rotation $\alpha$.}
\end{figure*}

To prepare the system probe, we perform beam expansion on the laser beam operating at $\SI{780}{\nm}$, followed by conversion to a $N$-order LG beam with the maximum OAM value $l=N$. Subsequently, a half-wave plate (HWP) and a quarter-wave plate (QWP) are utilized to initialize the polarization state as $\ang{45}$ polarized state $|+\rangle$. The indefinite rotation process of the light beam is implemented through a Sagnac interferometer, where the horizontally polarized component $|H\rangle$ propagates along the counterclockwise direction and the vertically polarized component $|V\rangle$ propagates along the clockwise direction. In the interferometer, a Dove prism is utilized to enable the angular rotation of beam profile. Here, we affix 4 piezoelectric transducer (PZT) chips onto the reflective surface of the Dove prism, arranged in a $2\times 2$ array. This configuration enable us to generate a tiny angular rotation signal for the beam profile.

Following the theoretical analysis in the last section, the angular rotation is finally transformed to the relative phase the $|H\rangle$ and $|V\rangle$ polarization components of the auxiliary meter. In practice, the interferometer inevitably introduces an additional relative phase $\Delta\varphi$ between the $|H\rangle$ and $|V\rangle$ polarization components. Considering this imperfection of the experimental system, there is a total relative phase $\Phi=2l\alpha+\Delta\varphi$ on the auxiliary meter after the indefinite rotation process. By performing the orthogonal projective measurements $\left\{|L\rangle\langle L|, |R\rangle\langle R|\right\}$ via a polarizing beam splitter (PBS) in combination with a HWP and a QWP, the relative phase $\Phi$ is able to be estimated directly from the measured optical powers of the two output ports of PBS:
\begin{align}
	\tilde{I}_{1} &= \frac{1}{2}I_{0}\left[1+\sin\left(2l\alpha+\Delta\varphi\right)\right], \label{eq:16} \\
	\tilde{I}_{2} &= \frac{1}{2}I_{0}\left[1-\sin\left(2l\alpha+\Delta\varphi\right)\right], \label{eq:17}
\end{align}
which leads to the estimator of relative phase
\begin{equation}
	\label{eq:18}
	\hat{\Phi} = \arcsin\left(\frac{\Delta\tilde{I}}{\Sigma\tilde{I}}\right) = \arcsin\left(\frac{\tilde{I}_{1}-\tilde{I}_{2}}{\tilde{I}_{1}+\tilde{I}_{2}}\right)
\end{equation}

In our experiments, the optical axis of the QWP is set to $\ang{45}$ from the horizontal plane, and the HWP is used to compensate the additional relative phase $\Delta\varphi$ to near $\ang{0}$. (The specific approach used for compensation is discussed in detail at the supplementary information.) First, we set the rotation angle $\alpha$ to $\ang{1}$ approximately via a manual adjuster of the Dove prism. Here, we record the experimental data of the measured optical powers from the two photodetectors using an oscilloscope with an acquisition time of $\SI{100}{\ms}$. To showcase the amplification of rotation angles with OAM values, we prepare Gaussian beam without vortex phase, as well as LG beams with $l=1, 4, 7, 10, 20, 30$ separately for individual experiments. In Fig. \ref{fig:5}, we illustrate the measured optical powers of PD1 and PD2 with blue lines and purple lines respectively. The unit of measurement is in mV as the trans-impedance amplifier of the photodetector converts the measured photocurrent into a voltage. Besides, we represent the differential optical powers of these two photodetectors with gray shadows. Then we estimate the corresponding relative phases and illustrate it in Fig. \ref{fig:5} with red lines.

\begin{figure}[htp]
	\centering
	\includegraphics[width=0.8\linewidth]{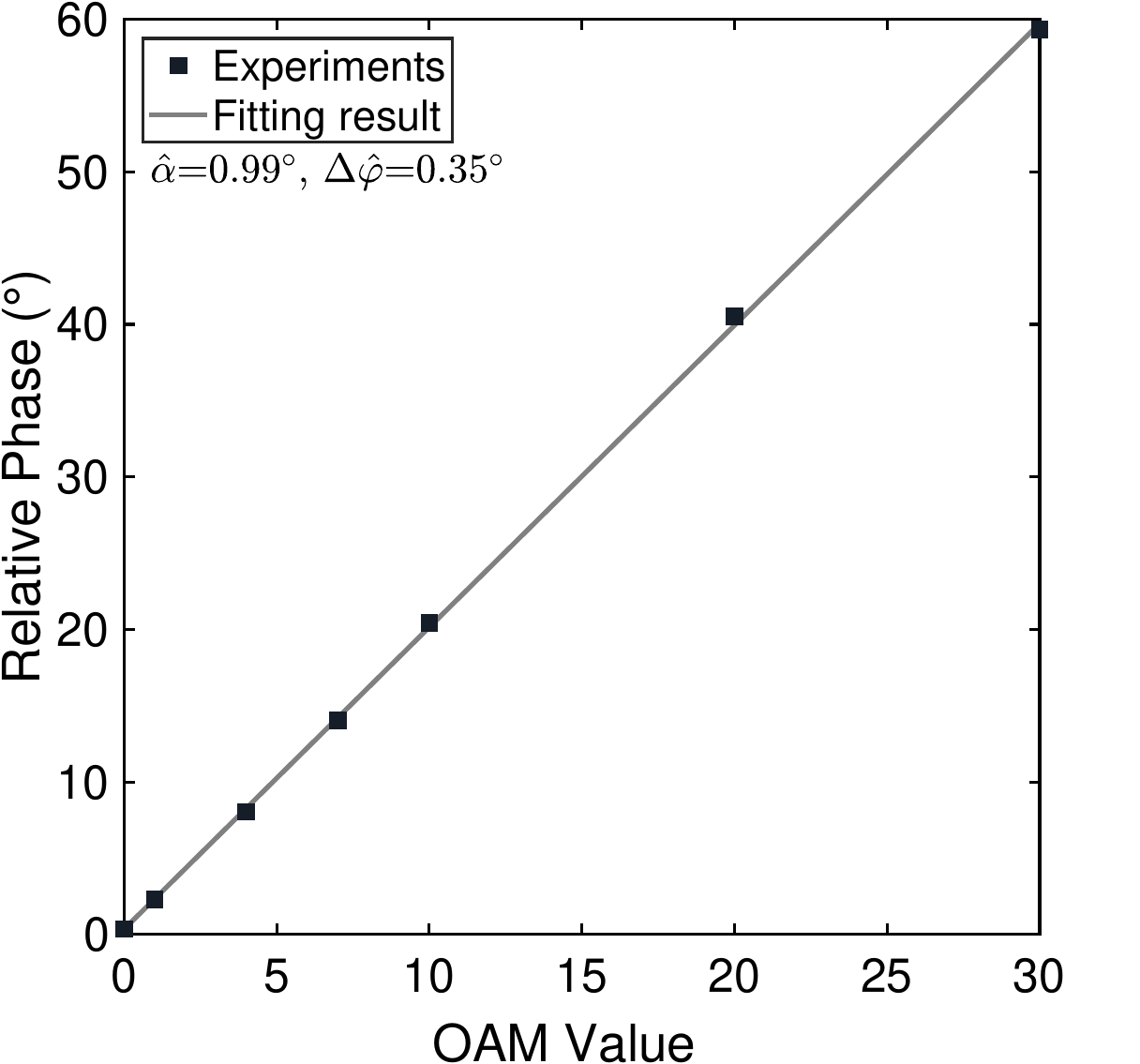}
	\caption{\label{fig:6} Fitting results of the linear increasing relations between the relative phase $\hat{\Phi}$ and the OAM value $l$. The black squares are the experimental results of demodulated relative phases with respect to OAM values 1, 4, 7, 10, 20, and 30. The gray line is the fitting result of the linear relation between the relative phase and the OAM value.}
\end{figure}

Fig. \ref{fig:5}\textbf{\textsf{a}} depicts the experimental results of inputting the Gaussian beam, where the relative phase $\Phi$ is solely determined by the additional relative phase $\Delta\varphi$ induced by systemic imperfections. Fig. \ref{fig:5}\textbf{\textsf{b}} through \ref{fig:5}\textbf{\textsf{g}} depict the experimental results of inputting $N$-order LG beams with OAM values of $l=N=1, 4, 7, 10, 20, 30$, respectively. As shown in Fig. \ref{fig:5}, the demodulated relative phase $\hat{\Phi}$ increases with the OAM values linearly. To evaluate the linear relationship between the relative phase and the OAM value, we fit the experimental results of demodulated relative phases $\hat{\Phi}$ at different inputting OAM values $l$ using the function $\hat{\Phi}=2l\hat{\alpha}+\Delta\hat{\varphi}$ in Fig. \ref{fig:6}.

As shown in Fig. \ref{fig:6}, we can demodulate the rotation angle $\hat{\alpha}=\ang{0.99}$ and the additional relative phase $\Delta\hat{\varphi}=\ang{0.35}$ from the fitting parameters. Notably, the coefficient of determination (R-square) for out fitting results is 99.98\%, indicating that the linear relationship between the experimental results of the relative phase $\hat{\Phi}$ and the OAM value $l$ is well explained by the function $\hat{\Phi}=2l\hat{\alpha}+\Delta\hat{\varphi}$.
\begin{figure}[htp]
	\centering
	\includegraphics[width=\linewidth]{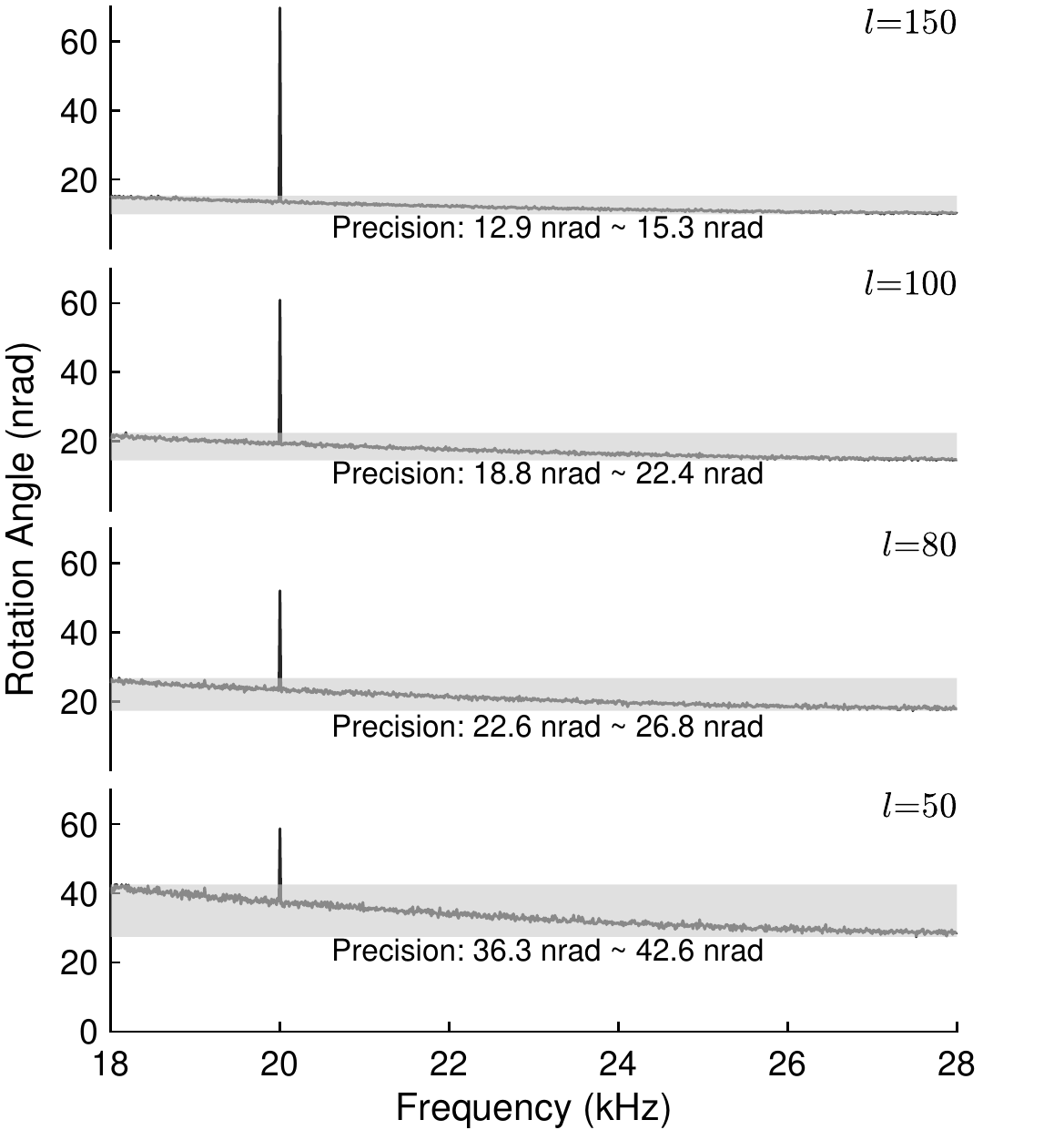}
	\caption{\label{fig:7} Amplitude spectrum of demodulated rotation angle $\alpha$ regarding to inputting $N$-order LG beams with OAM values of $l=N=50, 80, 100, 150$. Peak values at $\SI{20}{\kHz}$ correspond to the amplitudes of marking rotation signals. The gray shadows represent the noise floors at the frequency range of $\SI{18}{\kHz}$ to $\SI{28}{\kHz}$, which refer to the precisions of angular rotation measurement in the experiments.}
\end{figure}

In order to determine the ultimate precision limit on the measurement of angular rotations in our scheme, we employ the $N$-order LG beams with OAM values of $l=N=50, 80, 100, 150$. Here, we detect the optical power signal at an acquisition rate of $\SI{60}{kSa/\s}$, and demodulate the rotation angle $\alpha$. Then we calculate the corresponding amplitude spectrum from the demodulated signal with a time length of $\SI{0.1}{\s}$. We illustrate the results in Fig. \ref{fig:7}, where the frequency range is from $\SI{18}{\kHz}$ to $\SI{28}{\kHz}$.

As depicted in Fig. \ref{fig:7}, we introduce a sinusoidal signal of angular rotation at $\SI{20}{\kHz}$ as a marker, which is generated by applying two sinusoidal signals with opposite phases to the top-row and bottom-row PZT chips on the Dove prism. Here the peak-to-peak level of the sinusoidal signals applied to the PZT chips is $\SI{12}{\mV}$, which leads to tens of nanoradians of angular rotation of beam profile. In practice, the signals applied to the PZT chips also introduce an additional interference at the same frequency as the angular rotation signal due to the misalignment of the beam direction and the rotating axis of Dove prism. In the supplementary information, we have discussed how to demodulate the correct amplitude of the rotation angle in the presence of same-frequency interference, and evaluated the signal-to-interference ratio (SIR). From the noise floors of the corresponding amplitude spectra in Fig. \ref{fig:7}, we can obtain the precision limits for measuring angular rotations with OAM values of $l=50, 80, 100, 150$. As a result, the measurement precision is significantly improved by increasing the OAM value, and a precision of the nano-radian scale is achieved for the measuring of angular rotations via our OAM-polarization indefinite rotation process.

\section{Conclusion}
In summary, we have addressed the challenge of achieving nanoradian-scale precision on light rotation measurement by incorporating the strategy of indefinite time direction. Through the indefinite quantum dynamics in our scheme, we have maximized the utilization of resources in a quantum parameter estimation process. In experiment, we have devised an optical system that utilizes this quantum strategy to enhance the precision of light rotation measurements, resulting in the maximized utilization of OAM resources. Consequently, a remarkable precision of $\SI{12.9}{\nano rad}$ on light rotation measurement has been achieved with assistance of 150-order LG beam. Furthermore, as the IQPE scheme has theoretically exhibited an enhanced quantum limit within spin systems, it holds significant implications for various quantum metrological systems, including nitrogen vacancy (NV) spin qubits\cite{doi:10.1146/annurev-physchem-040513-103659,doi:10.1021/acssensors.1c00415} and nuclear magnetic resonance (NMR) sensors\cite{doi:10.1021/acs.chemrev.8b00202,D2CC01546C}.

\section{Materials and methods}
\subsection{Experimental materials}
We utilized a single-frequency fiber laser, specifically the NKT Photonics Koheras HARMONIK series, operating at a wavelength of $\SI{780}{\nm}$. The SLM used in our setup is the Hamamatsu Photonics X13138-02, featuring a resolution of $1272\times 1024$ pixels with a pixel pitch of $\SI{12.5}{\um}$. The SLM is employed to generate the LG beams, and the detailed generation method can be found in the supplementary information. The 4-f system includes two Fourier lenses with a focal length of $\SI{25}{\cm}$. Additionally, we incorporate an iris diaphragm as a spatial filter in our experimental configuration. The PZT chips used for actuating the Dove prism are sourced from Core Tomorrow Company (part number: NAC2013). These PZT chips shift approximately $\SI{22}{\nm}$ when driven by a voltage of $\SI{1}{\V}$. The interval between the top-row and bottom-row PZT chips is approximately $\SI{10}{\mm}$. Hence, applying a driving signal with a peak-to-peak level of $\SI{12}{\mV}$ results in a rotation amplitude of approximately $\SI{60}{\nano rad}$. Our experimental results in Fig. \ref{fig:7} align well with this estimating value. For the projective measurements in our experiment, we utilized photodetectors from Thorlabs Inc. (part number: PDA100A2) to receive the projective optical powers. These photodetectors have a responsivity of $\SI{0.585}{\A/\W}$ at a wavelength of $\SI{780}{\nm}$. In our setup, the transresistance gain was set to $\SI{15}{\kV/\A}$.

\subsection{Upper bound of quantum Fisher information}
For an arbitrary pure quantum state $|\psi(g)\rangle$, where $g$ is an unknown parameter encoded in this state, its corresponding QFI is defined as\cite{DEMKOWICZDOBRZANSKI2015345,Liu_2019}
\begin{equation}
	\label{eq:19}
	\mathcal{Q}(g) = 4\left(\frac{\partial\langle\psi(g)|}{\partial g}\frac{\partial|\psi(g)\rangle}{\partial g}-\left|\frac{\partial\langle\psi(g)|}{\partial g}|\psi(g)\rangle\right|^{2}\right).
\end{equation}
Substituting final state $|\psi(g)\rangle=\hat{U}_{S}(0\to T)|\psi_{i}\rangle$ into this equation, it is easy to obtain the QFI in the SQPE scenario as $\mathcal{Q}_{S}(g)=4\mathrm{Var}[\hat{\mathcal{H}}_{S}(g)]_{|\psi_{i}\rangle}$, where $\hat{\mathcal{H}}_{S}(g)=\mathrm{i}\hat{U}^{\dagger}_{S}(0\to T)\partial_{g}\hat{U}_{S}(0\to T)$ is the generator. By rewritten the generator as $\hat{\mathcal{H}}_{S}(g)=\int_{0}^{T}\hat{U}_{S}^{\dagger}(0\to t)\hat{V}_{S}\hat{U}_{S}(0\to t)\mathrm{d}t$, where $\hat{V}_{S}=\partial_{g}\hat{H}_{S}$ is the characteristic operator of the parameterizing dynamics. Since the maximum and minimum eigenvalues of $\hat{V}_{S}$ are denoted as $\lambda_{M}(t)$ and $\lambda_{m}(t)$, separately, the the maximum and minimum eigenvalues of $\hat{\mathcal{H}}_{S}$ can be derived as $\int_{0}^{T}\lambda_{M}(t)\mathrm{d}t$ and $\int_{0}^{T}\lambda_{m}(t)\mathrm{d}t$. Hence, the upper bound of the QFI in the SQPE scheme can be calculated as
\begin{equation}
	\label{eq:20}
	\mathcal{Q}_{S}(g) = 4\mathrm{Var}\left[\hat{\mathcal{H}}_{S}(g)\right]_{|\psi_{i}\rangle} \le \left\{\int_{0}^{T}\left[\lambda_{M}(t)-\lambda_{m}(t)\right]\mathrm{d}t\right\}^{2}.
\end{equation}

Similarly, the QFI in our IQPE scenario can be obtained as $\mathcal{Q}_{I}(g)=4\mathrm{Var}[\hat{\mathcal{H}}_{I}(g)]_{|\psi_{i}\rangle|\phi_{i}\rangle}$, where the corresponding generator is given by
\begin{equation}
	\label{eq:21}
	\hat{\mathcal{H}}_{I}(g) = \hat{\mathcal{H}}_{S}|0\rangle\langle 0|-\hat{U}_{S}\hat{\mathcal{H}}_{S}\hat{U}_{S}^{\dagger}|1\rangle\langle 1| = \left(
	\begin{array}{cc}
		\hat{\mathcal{H}}_{S} & 0 \\
		0 & -\hat{U}_{S}\hat{\mathcal{H}}_{S}\hat{U}_{S}^{\dagger}
	\end{array}
	\right).
\end{equation}
It is easy to derive that the maximum eigenvalue of $\hat{\mathcal{H}}_{I}(g)$ is $\max[|\int_{0}^{T}\lambda_{M}(t)\mathrm{d}t|,|\int_{0}^{T}\lambda_{m}(t)\mathrm{d}t|]$, and the minimum eigenvalue is $\min[-|\int_{0}^{T}\lambda_{M}(t)\mathrm{d}t|,-|\int_{0}^{T}\lambda_{m}(t)\mathrm{d}t|]$. Thus, the upper bound of the QFI in the SQPE scheme can be calculated as
\begin{equation}
	\label{eq:22}
	\mathcal{Q}_{I}(g) = 4\mathrm{Var}\left[\hat{\mathcal{H}}_{I}(g)\right]_{|\psi_{i}\rangle|\phi_{i}\rangle} \le \max\left\{\left[\int_{0}^{T}2\lambda_{M}(t)\mathrm{d}t\right]^{2}, \left[\int_{0}^{T}2\lambda_{m}(t)\mathrm{d}t\right]^{2}\right\}.
\end{equation}

\subsection{Geometric representations of polarization states and HLG beams}
In the main text, we have represented the polarization states on a classical PS, as shown in Fig. \ref{fig:2}\textbf{\textsf{a}}. Practically, the axes $\hat{S}_{1}$, $\hat{S}_{2}$ and $\hat{S}_{3}$ represent the Stokes parameters, which can be described as $\hat{S}_{1}=|H\rangle\langle H|-|V\rangle\langle V|$, $\hat{S}_{2}=|+\rangle\langle +|-|-\rangle\langle -|$ and $\hat{S}_{3}=|R\rangle\langle R|-|L\rangle\langle L|$. By assigning the circular polarization states $|R\rangle$ and $|L\rangle$ as the orthogonal bases, the operators $\hat{S}_{1}$, $\hat{S}_{2}$ and $\hat{S}_{3}$ can be determined by the Pauli matrices
\begin{equation}
	\label{eq:23}
	\hat{S}_{1} = \left(
	\begin{array}{cc}
		0 & 1 \\
		1 & 0
	\end{array}
	\right), \;
	\hat{S}_{2} = \left(
	\begin{array}{cc}
		0 & -\mathrm{i} \\
		\mathrm{i} & 0
	\end{array}
	\right), \;
	\hat{S}_{3} = \left(
	\begin{array}{cc}
		1 & 0 \\
		0 & -1
	\end{array}
	\right).
\end{equation}
Then an arbitrary polarization state can be obtained from the right-handed circular polarization state $|R\rangle$ (which is at the northern pole of the classical PS) through an Euler rotation
\begin{align}
	|\psi_{i}\rangle &= \exp(-\mathrm{i}\frac{\hat{S}_{3}}{2}\phi)\exp(-\mathrm{i}\frac{\hat{S}_{2}}{2}\theta)|R\rangle \nonumber \\
	&= \cos(\frac{\theta}{2})|R\rangle+\sin(\frac{\theta}{2})\mathrm{e}^{\mathrm{i}\phi}|L\rangle \label{eq:24}
\end{align}

Next, utilizing the analogy between the HLG modes and the quantum eigenstates of a 2DHO, the axes $\hat{J}_{1}$, $\hat{J}_{2}$ and $\hat{J}_{3}$ of the modal PS can be determined using the Schwinger oscillator model
\begin{align}
	\hat{J}_{1} &= \frac{1}{2w^{2}}\left(\hat{x}^{2}-\hat{y}^{2}\right)+\frac{w^{2}}{8}\left(\hat{p}_{x}^{2}-\hat{p}_{y}^{2}\right), \nonumber \\
	\hat{J}_{2} &= \frac{1}{w^{2}}\hat{x}\hat{y}+\frac{w^{2}}{4}\hat{p}_{x}\hat{p}_{y}, \quad \hat{J}_{3} = \frac{1}{2}\left(\hat{x}\hat{p}_{y}-\hat{y}\hat{p}_{x}\right), \label{eq:25}
\end{align}
where, in the position representation, $\hat{x}\mapsto x$ and $\hat{p}_{x}\mapsto-\mathrm{i}\partial_{x}$ (and similarly to $y$). These operators satisfy the commutation relations of algebra $\mathfrak{su}(2)$, $[\hat{J}_{a},\hat{J}_{b}]=\mathrm{i}\sum_{c}\epsilon_{abc}\hat{J}_{c}$, with $a$, $b$, $c$ = 1, 2, or 3 and $\epsilon_{abc}$ being the Levi-Civita tensor. As shown in Fig. \ref{fig:2}\textbf{\textsf{d}}, the poles represent the LG modes with opposite OAM values. The field distribution of the LG beam at the focal plane is given by
\begin{align}
	\label{eq:26}
	\psi_{p,l}^{\mathrm{LG}}(\mathbf{r}) =& \frac{1}{w}\sqrt{\frac{2^{|l|+1}p!}{\pi\left(p+|l|\right)!}}L_{p}^{|l|}\left(\frac{2r^{2}}{w^{2}}\right) \nonumber \\
	&\times\left(\frac{r}{w}\right)^{|l|}\exp\left(-\mathrm{i}l\varphi\right)\exp\left(-\frac{r^{2}}{w^{2}}\right),
\end{align}
where $L_{p}^{|l|}(\cdot)$ is the generalized Laguerre polynomial, $p$ is the radial index and $l$ is the topological charge (which determines OAM value) of LG beam. The total order is denoted as $N=2p+|l|$, which determines the Gouy Phase. Then the LG beam can be denoted by the ket $|N,l\rangle$. Drawing on the analogy between the HLG modes and 2DHO model, the indices $N$ and $l$ also serve as the spin quantum numbers, and the orthogonal kets $\{|N,l\rangle|l=-N,-N+2,\dots,N-2,N\}$ form the $(N+1)$-dimensional representation space of the SU(2) group. Since the SU(2) group is isomorphic to the SO(3) group, operators $\hat{J}_{1}$, $\hat{J}_{2}$, and $\hat{J}_{3}$ serve as the rotation generators in the corresponding SO(3) group. Consequently, an arbitrary HLG beam state of the same order can be obtained from LG beam $|N,l\rangle$ on the modal PS through an Euler rotation
\begin{equation}
	\label{eq:27}
	|\psi_{i}\rangle = \exp(-\mathrm{i}\hat{J}_{3}\phi)\exp(-\mathrm{i}\hat{J}_{2}\theta)|N,l\rangle.
\end{equation}
In practical optical systems, the rotation transformation along the $\hat{J}_{3}$-axis can be achieved by using a Dove prism (rotation of beam profile). The rotation transformation along the $\hat{J}_{1}$-axis can be achieved by using a pair of cylindrical lenses (astigmatic transformation). The rotation transformation along $\hat{J}_{2}$-axis can be realized by combining rotations along the $\hat{J}_{3}$- and $\hat{J}_{1}$-axes.

\subsection{Ultimate precision of angular rotations}
As the parameterized state of the entire system after the indefinite rotation process is $|\Psi(\alpha)\rangle = \frac{1}{\sqrt{2}}(\mathrm{e}^{-\mathrm{i}l\alpha}|H\rangle+\mathrm{e}^{\mathrm{i}l\alpha}|V\rangle)|l\rangle$. By applying a pair of the projective measurements $\left\{|L\rangle\langle L|, |R\rangle\langle R|\right\}$ on the auxiliary meter, the probabilities of the orthogonal projections onto the measurement outcomes are given by:
\begin{align}
	P_{L} &= \frac{1}{2}\left[1+\sin\left(2l\alpha\right)\right], \label{eq:28}\\
	P_{R} &= \frac{1}{2}\left[1-\sin\left(2l\alpha\right)\right]. \label{eq:29}
\end{align}
Based on the classical estimation theory, the classical Fisher information (CFI) of rotation parameter $\alpha$ is calculated as $\mathcal{F}(\alpha)=4l^{2}=4N^{2}$. Drawing on the classical Cramér-Rao bound theory, the ultimate precision of the rotation parameter $\alpha$ can be expressed as shown in Equation (\ref{eq:15}), using the inequality $\delta\hat{\alpha}^{2}\ge 1/[\nu\mathcal{F}(\alpha)]$. Furthermore, the theoretical optimal estimator is given as
\begin{equation}
	\label{eq:30}
	\hat{\alpha} = \frac{1}{2l}\arcsin\left(\frac{\Delta\tilde{\nu}}{\Sigma\tilde{\nu}}\right) = \frac{1}{2l}\arcsin\left(\frac{\Delta\tilde{I}}{\Sigma\tilde{I}}\right)
\end{equation}
where $\Delta\tilde{\nu}=\tilde{\nu}_{L}-\tilde{\nu}_{R}$ is the difference in detected photon numbers between the projective bases $|L\rangle$ and $|R\rangle$, while $\Sigma\tilde{\nu}=\tilde{\nu}_{L}+\tilde{\nu}_{R}$ is their sum. In practice, the measured optical power is proportional to the number of photons detected by the photodiode (PD). Consequently, the rotation angle $\alpha$ can be estimated by calculating the differential optical power $\Delta\tilde{I}$ and the total optical power $\Sigma\tilde{I}$ with respect to the projective bases $|L\rangle$ and $|R\rangle$.

\begin{acknowledgments}
	This work is financially supported by the National	Natural Science Foundation of China (Grant No. 62071298) and the Innovation Program for Quantum Science and Technology (Grant No. 2021ZD0300703).
\end{acknowledgments}

\bibliography{bibliography}

\end{document}